\documentclass[Crown,sagev,times, doublespace]{sagej}
\usepackage{comment}
\usepackage[revision]{revdiff} %make 'clean' ipv 'revision' for no notes!
\usepackage{moreverb,url}
\usepackage{bm, mathtools}
\usepackage{threeparttablex}
\usepackage{booktabs}

\usepackage[colorlinks,bookmarksopen,bookmarksnumbered,citecolor=red,urlcolor=red]{hyperref}

\newcommand\BibTeX{{\rmfamily B\kern-.05em \textsc{i\kern-.025em b}\kern-.08em
T\kern-.1667em\lower.7ex\hbox{E}\kern-.125emX}}

\def\volumeyear{2020}

\begin{document}

\runninghead{Zhan et al.}

\title{Joint modeling with time-dependent treatment and heteroscedasticity: Bayesian analysis with application to the Framingham Heart Study}

\author{ Sandra P. Keizer\affilnum{1,*}, Zhuozhao Zhan\affilnum{1,*}, Vasan S. Ramachandran\affilnum{2} and Edwin R. van den Heuvel\affilnum{1,2}}

\affiliation{\affilnum{1}Department of Mathematics and Computer Science,
Eindhoven University of Technology, The Netherlands\\
\affilnum{2} Center of Integrative Transdisciplinary Epidemiology, Preventive Medicine \& Epidemiology, Boston University School of Medicine, USA\\
\affilnum{*} These authors contributed equally.
}

\corrauth{Sandra P. Keizer, Department of Mathematics and Computer Science,
Eindhoven University of Technology, The Netherlands}

\email{s.p.keizer@tue.nl}

\begin{abstract}
Medical studies for chronic disease are often interested in the relation between longitudinal risk factor profiles and the risk of disease outcomes in later life. These profiles may be subject to intermediate structural changes due to treatment or environmental influences. Analysis of such studies may be handled by the joint model framework. However, current joint modeling does not consider either structural changes in the residual variability of the risk profile or the influence of subject-specific residual variability of the risk profile on the time-to-event outcome. In the present paper, we extend the joint model framework to address these two heterogeneous intra-individual variabilities. A Bayesian approach is used to estimate the unknown parameters and simulation studies are conducted to investigate the performance of the method. The proposed joint model is applied to data from the Framingham Heart Study to investigate the influence of anti-hypertensive medications on the systolic blood pressure variability together with the effect of such medication use on the risk of developing cardiovascular disease. We show that anti-hypertensive medication use is associated with elevated systolic blood pressure variability and such increased systolic pressure variability is associated with an increased risk of developing cardiovascular disease.
\end{abstract}


\maketitle

\section{Introduction}\label{sec:intro}
Medical studies on the risk of chronic diseases, such as cancer, human immunodeficiency virus (HIV), and cardiovascular diseases (CVD), frequently involve studying longitudinal measurements of certain risk factors in relation to individuals' later life risk of disease outcomes. The trajectories, or the longitudinal profiles, of the risk factors which describe the progression of the risk factor variables over time, may play an important role in preventive healthcare. Indeed, the risk factors are commonly monitored to make decision about timely interventions and to help predict later life diseases. Thus, in observational studies, the longitudinal profiles of risk factors are subject typically to intermediate structural changes caused by factors such as treatment interventions, behavioral adaptations, or environmental changes. Disruptions of the longitudinal profile typically may have an effect on the event outcome. Furthermore, structural changes in the longitudinal profile of risk factors may include immediate changes in the values for the risk factor, changes in growth rate of the risk profile trajectory, and changes in the variability of the risk factor. All these changes may happen simultaneously and they may have distinctive effects on the disease outcomes. These phenomena are commonly observed in life course epidemiology~\cite{Kuh2002, Kuh2004}.

To address research questions about the association of the longitudinal profiles of risk factors and the time to outcome events of interest, statistical analyses are often conducted with some form of joint models. These joint models often assume either the existence of a set of low-dimensional unobserved time-independent variables that vary across individuals and operate underneath both the longitudinal profile and the event time~\cite{Wulfsohn1997}, or that the risk of occurrence of an outcome event is directly associated with the latent (functions of) smooth longitudinal profile~\cite{Rizopoulos2012,Papageorgiou2019}. Joint models focus on how the levels of the smooth profiles of risk factors are associated with prognosis~\cite{Tsiatis2004}. Recent advances in joint models extended single longitudinal risk factor profiles to multiple risk factor profiles with possible extension to different risk factor types~\cite{Larsen2004, Chi2006, Andrinopoulou2014, Musoro2015, Proust2014}. Other recent advances in joint model include the incorporation of heteroscedasticity  in the longitudinal profile\cite{Barret2018} or the inclusion of a random change point in the longitudinal profile\cite{Yu2010}.
Nevertheless, the main associations between the longitudinal risk factor profiles and the time-to-event outcomes still closely follow the two approaches described above. Whilst, relations between other features of the longitudinal profiles, such as their intra-individual variability and the time-to-event outcome is less frequently investigated. 

The present investigation is motivated by data from the Framingham Heart Study~\cite{dawber1951}, an observational cohort study across multiple generations aimed to identify the common factors or characteristics that contribute to CVD. In the past, the Framingham Heart Study data have been used to investigate the associations between various blood pressure components, including systolic blood pressure (SBP), diastolic blood pressure (DBP), and pulse pressure (PP), with the risk of developing CVD (see for example  Levy~\cite{levy1999}; Kannel~\cite{kannel1996}; Franklin~\cite{Franklin2015}; Nayor et al.~\cite{Nayor2018}). In addition, the Framingham Heart Study data have been used to study the hemodynamic patterns of age-related changes in blood pressure~\cite{Kannel1971, Franklin1999, Franklin2009}. Hathaway and D’Agostino~\cite{Hathaway1993} reported a significant association between variance of SBP and subsequent coronary heart disease among a group of 516 women in the Framingham Heart Study. Their approach resembled a two-step joint model (an earlier version of the current joint model framework). Summary statistics, such as the mean, the variance or average difference, based on regression of the SBP on age and the original observed repeated measurements of SBP were considered as independent predictors in a logistic regression model for coronary heart disease. This two-step approach, however, is known to produce biased estimates in the joint model literature~\cite{Tsiatis2004, Rizopoulos2012}.

It has been established that increased long-term SBP variability is associated with an increased risk of CVD events and mortality~\cite{Stevens2016,Sponholtz2019}. However, currently available methods for evaluating SBP variability are diverse - they include standard deviation, coefficient of variation, average real variability~\cite{Mena2005} (sum of absolute difference between two consecutive SBPs), and variability independent of mean~\cite{Rothwell2010}. Several methodological issues have been raised related to these analysis and our understanding of SBP variability~\cite{Stevens2016}. First, analysis of variability need to take into account the correlation between high mean blood pressure and high variability. Second, variabilities based on the repeated measurements during follow-up are frequently used in studies as a baseline risk factor, which potentially introduces immortal-time bias and other related problems that joint models attempts to address. 

In the present investigation, we extend the commonly used joint model~\cite{Rizopoulos2012} for a longitudinal profile with heterogeneous intra-individual variabilities and relate the risk of developing chronic disease to the intra-individual variability in combination with disruptive risk profiles due to treatment interventions. A partial likelihood of the aforementioned statistical model is proposed that is proportional to the complete likelihood where the initiation of treatment of a risk factor is assumed to be associated with the observed value of the longitudinal profile of the risk factor but does not share parameters with the longitudinal and time-to-event models. A Monte Carlo Markov Chain (MCMC) algorithm was used to estimate the parameters in the longitudinal and the time-to-event models.

The rest of this paper is organized as follows. In Section~\ref{sec:method}, we propose a joint model for the longitudinal profile, the treatment initiation for a given risk factor, and the event outcome of interest. Furthermore, we construct the likelihood function for the proposed model, which leads to the partial likelihood function for the longitudinal and time-to-event outcome only. We provide the prior distributions for the parameters of interests and the posterior sampling scheme in Section~\ref{sec:mcmc}. In Section~\ref{sec:sim}, a simulation study is described. In Section~\ref{sec:fhs}, the proposed joint model is fitted to the Framingham Heart Study data and the effect of the SBP variability on the risk of developing CVD and the effect of the anti-hypertensive medication on the SBP variability are estimated. Discussion follows in Section~\ref{sec:discussion}.

\section{Motivating example}\label{sec: fhs_eda}
Data of the first generation cohort of the Framingham Heart Study is considered in this investigation. The Framingham Heart Study started in 1948 with its initial enrollment of an original cohort of 5209 men and women between the ages of 30 and 62 years from the town of Framingham, Massachusetts, United States of America. The collected data consisted of extensive medical history, physical examinations performed independently by at least two physicians, chest X-ray, electrocardiogram, and blood and urine samples. We focus on SBP, measured every two years during follow-up together with the status on whether anti-hypertensive medications were administrated since the last examination. Furthermore, CVD status at each examinations was also recorded. The interest is to investigate (1) the effect of the anti-hypertensive medications on the SBP, (2) its related risk of developing CVD, and (3) the effect of SBP variability on the risk of developing CVD. 

Preliminary analysis of the Framingham Heart Study data was conducted by fitting the subject-specific growth model~\eqref{eq:long}: 
\begin{equation*}
\begin{cases}
    &P(Y_{ij} \le y_{ij} | \bm{r}_i,Z_{i,j-1}) = \Phi\left(\frac{y_{ij}-\mu_{ij}}{\sigma_{0}}\right),\\
    &\mu_{ij}=b_{0i}+b_{1i}t_{ij}+Z_{i,j-1}\left\{b_{2i}+b_{3i}(t_{ij}-s_i)\right\}
\end{cases}
\end{equation*}

\begin{figure}[htp]
\centering
\includegraphics[width=0.7\textwidth]{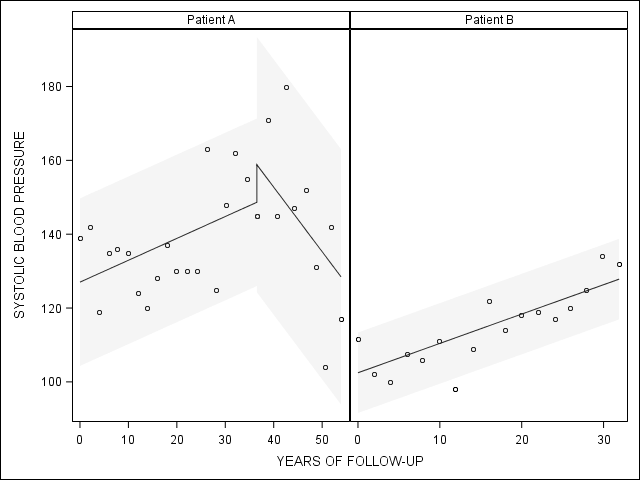}
\caption{Systolic blood pressure of two individuals fitted with a homoskedastic subject-specific model. Dots: Original observations; Lines: Conditional mean profile: $\hat{\mu}_{ij}$; Shaded area: 95\% CI: $\hat{\mu}_{ij}\pm 1.96 \mathrm{SD}(y_{ij}-\hat{\mu}_{ij})$ (SD: standard deviation before and after treatment)}\label{fig:patients}
\end{figure}

\begin{figure}[htp]
\centering
\includegraphics[width=0.8\textwidth]{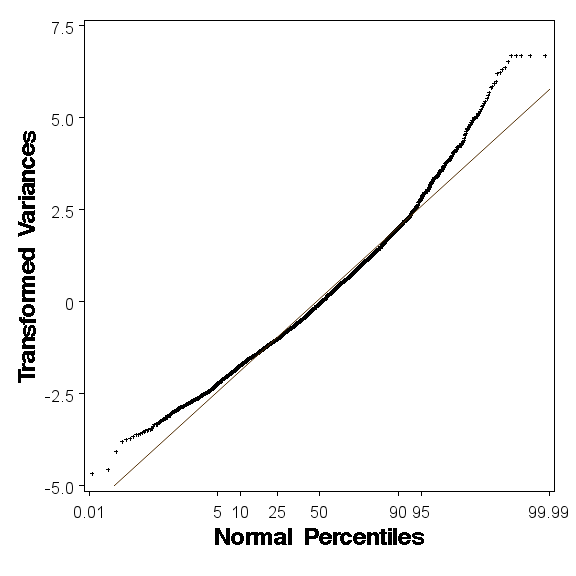}
\caption{Normality plot of the transformed residual variances of individuals before treatment}\label{fig:pp}
\end{figure}

\begin{figure}[htp]
\centering
\includegraphics[width=0.8\textwidth]{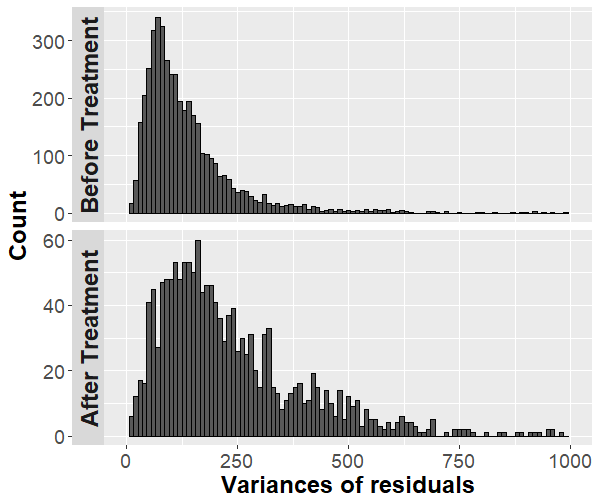}
\caption{Residual variances of individuals on systolic blood pressure before and after treatment.}\label{fig:hist}
\end{figure}

Results indicate that the residual variance across individuals for SBP before treatment is heteroscedastic. Figure~\ref{fig:patients} displays that the SBP variability before treatment differs between two individuals. In addition, it can be shown that the estimated residual variance for SBP before treatment does not follow a chi-square distribution. Under homoscedasticity, it was expected that the sum of squares of the $q_i$ residuals $(q_i - 1)s_{0i}^2$ for individual $i$ before treatment, divided by the variance of all residuals before treatment $\sigma_0^2$, has a chi-square distribution with $q_i-1$ degrees of freedom:
\begin{equation*}
    \frac{1}{\sigma_0^2}(q_i-1) s_{0i}^2 \sim \chi^2_{q_i - 1}.
\end{equation*} 
This distributional assumption remains approximately true for large numbers of participants when $\sigma_0^2$ is replaced by its estimate $\hat{\sigma}_0^2$ over all residuals before treatment. However, Figure~\ref{fig:pp} demonstrates a clear violation of normality for the transformation $\Phi^{-1}\left(\chi_{q_i-1}^2 ((q_i-1)s_{0i}^2/\hat{\sigma}_0^2)\right)$ (Anderson-Darling $A^2=16.11$, $p<0.005$). Clearly, the variability in the variances exceeds the variance of the chi-square distribution. Another important observation is that the intra-individual variability changes after treatment within patients. The residual variance seems to increase which is depicted in Figure~\ref{fig:hist} where the histogram of the residual variances for SBP before and after treatment are depicted.

\section{Methods}\label{sec:method}

\subsection{Treatment and longitudinal profile}
Let $\bm{Y}'_i=(Y_{i1},\dots, Y_{im_i})$ denote the vector of repeated measurements of the longitudinal risk factor profile for individual $i\,(i=1,\dots,N)$ at measurement occasions $t_{ij}$ with $j=0,1,\dots, m_i$ and $t_{i0}=0$ indicating the baseline. For the corresponding treatment indicator, $\bm{Z}'_i=(Z_{i1},\dots,Z_{im_i})$, treatment is administrated at the $j$th measurement occasion if $Z_{ij}=1$, $Z_{ij}$ is $0$ otherwise. Assuming that treatment is administrated based on the observed value of the longitudinal profile, as well as the previous treatment condition of the individual, we may assume that 
\begin{equation}
    \label{eq:trt}
    P(Z_{ij}=1|Y_{ij}=y_{ij}; Z_{i,j-1})=Z_{i,j-1}+(1-Z_{i,j-1}) \Phi(\alpha_0+\alpha_1 y_{ij}),
\end{equation}
where $\Phi$ is the standard normal cumulative distribution function and $\bm{\alpha}=(\alpha_0, \alpha_1)'$ are unknown parameters related to the treatment process. Implicitly, we assume that $Y_{ij}$ precedes $Z_{ij}$, namely the observed value of $Y_{ij}$ is used to determine treatment administration at measurement $j$. In this treatment-model, the only factors that determines the initiation of treatment is the most recent measurement of the observed longitudinal profile and previous treatment condition. This may be a simplification of a real-life decision processes, since other measured factors could contribute to the treatment decision, but our treatment-model can be extended to include other factors, for example all of the past measurements of the longitudinal profile and treatment conditions. It should be noted that, when an individual is already under treatment $Z_{i,j-1} =1$, treatment will not stop, namely $P(Z_{ij}=1 | Y_{ij}, Z_{i,j-1}=1) =1$ as implied by the formulation of model~\eqref{eq:trt}. This may be realistic when medication is used for chronic conditions. Thus we will not consider the possibility of treatment cessation in the present investigation.

For the longitudinal profile, a subject-specific linear growth model is specified as
\begin{equation}
\label{eq:long}
\begin{cases}
    &P(Y_{ij} \le y_{ij} | \bm{r}_i,Z_{i,j-1}) = \Phi\left(\frac{y_{ij}-\mu_{ij}}{\sigma_{ij}}\right),\\
    &\mu_{ij}=b_{0i}+b_{1i}t_{ij}+Z_{i,j-1}\left\{b_{2i}+b_{3i}(t_{ij}-s_i)\right\},\\
    &\sigma_{ij}^2=\sigma_0^2\exp(\nu Z_{i,j-1}+c_i),
\end{cases}
\end{equation}
where the random term $\bm{r}_i=(b_{0i},b_{1i},b_{2i},b_{3i},c_i)'$ is assumed to follow a multivariate normal distribution $\bm{r}_i \sim \mathcal{N}(\bm{\theta}, \bm{\Sigma})$ with $\bm{\theta}=(\beta_0, \beta_1, \beta_2, \beta_3, 0)'$, $\sigma_0^2$ denotes the mean residual variance before treatment over all individuals, $t_{ij}$ is the time point of the $j$th measurement occasion for individual $i$, $s_i=t_{i,j_i}$ with $j_i=\inf\{j: Z_{ij}-Z_{i,j-1}=1\}$ the time when treatment is first initiated ($s_i$ is considered equal to $t_{i, m_i}$ if no treatment is administrated to individual $i$ during the follow-up). The intra-individual variance, $\sigma_{ij}^2$,  depends on the latent variable $c_i$ and treatment indicator $Z_{i,j-1}$ with a multiplicative treatment effect equal to $\exp(\nu)$ on the variance. The conditional mean $\mu_{ij}$ of the longitudinal profile is formulated such that treatment would have effects on both the absolute value and the rate of changes of the longitudinal profile. The four latent variables $b_{1i}, b_{2i}, b_{3i}, b_{4i}$ represents the subject-specific baseline profile, growth rate without treatment,immediate treatment effect on the profile, and change in growth rate after treatment, respectively. Furthermore, treatment is assumed to have a multiplicative effect that is common across individuals on the variability of the profile while the between-individual variability of the variance is captured by the latent variable $c_i$. The covariance matrix $\bm{\Sigma}$ allows the latent variable $c_i$ to be correlated with the latent variables $b_{1i}, b_{2i}, b_{3i}, b_{4i}$ in the conditional mean. This provides opportunities to investigate the correlation between the variabilities and different aspects of the longitudinal profile and understanding of how the individual variability is related to the longitudinal profile before and after treatment.

\subsection{Time to Event}
Let $T_i=\min(T_i^*,C_i)$ be the observed event time for individual $i$ taking the value of the true event time $T_i^*$ if the event is observed and otherwise taking the value of the censoring time $C_i$. The hazard rate $\lambda_i(t)$ is specified as
\begin{equation}\label{eq:surv}
    \lambda_i(t) = \lambda_0(t)\exp\left(\gamma_0 \mu_i(t)+ \gamma_1 \log\sigma_i^2(t)\right),
\end{equation}  
where $\mu_i(t) = b_{0i} + b_{1i} t + Z_{i,j-1}\{b_{2i} + b_{3i}(t - s_i)\}$, and $\sigma_i^2 (t) = \sigma_0^2\exp(\nu Z_{i,j-1}+c_i)$ with $\{j: t_{i,j-1} < t \le t_{ij}\}$, and $\gamma_0, \gamma_1$ their corresponding effects on the hazard function, and $\lambda_0(t)$ the baseline hazard function of a parametric survival distribution such as the Weibull distribution. In addition, the event indicator is denoted by $D_i = \bm{1}(\{T_i^*\le C_i\})$. The time-to-event model implies that the hazard ratio between two individuals $i$ and $j$ with the same conditional mean profile but different variabilities would be equal to
\begin{equation*}
    \frac{\lambda_i(t)}{\lambda_j(t)} =\frac{\exp(\gamma_1 \log \sigma_i^2)}{\exp(\gamma_1 \log \sigma_j^2)} = \exp\left\{\gamma_1 \log \left(\frac{\sigma_i^2}{\sigma_j^2}\right)\right\}.
\end{equation*}
Furthermore, no direct treatment effect is specified in the time-to-event model. This reflects the modeling assumption that the effect of the treatment is completely being mediated by the structural changes in the longitudinal profile. Alternatively, we can also incorporate a direct effect of the treatment into the hazard function. Different covariates can also be added into the hazard function.

\subsection{The likelihood function}
The joint distribution of the longitudinal pair $(\bm{Y}_i, \bm{Z}_i)$ and the event-time pair $(T_i, D_i)$ for individual $i$ is
\begin{equation*}
    f(\bm{Y}_i,\bm{Z}_i,T_i,D_i)=\int_{\mathbb{R}^5} f(\bm{Y}_i, \bm{Z}_i, T_i, D_i|\bm{r}_i)f(\bm{r}_i)d\bm{r}_i.
\end{equation*}
We will further assume, that conditional on the latent variables $\bm{r}_i$, the time-to-event outcome is independent of the longitudinal profile. As a consequence, the joint distribution can be expressed as
\begin{equation*}
\begin{split}
    f(\bm{Y}_i,\bm{Z}_i,T_i,D_i) &= f(Z_{i1}|Y_{i,1})\prod_{j=2}^{m_i}f(Z_{ij}|Y_{ij},Z_{i,j-1})\\
    &\int f(Y_{i1}|\bm{r}_i)\prod_{j=2}^{m_i} f(Y_{ij}|Z_{i,j-1},\bm{r}_i)\times f(T_i,D_i|\bm{r}_i) f(\bm{r}_i)d\bm{r}_i   
\end{split}
\end{equation*}
Since the treatment model~\eqref{eq:trt} does not depend on the latent variables $\bm{r}_i$, the term $f(Z_{ij}|Y_{i,j-1}, Z_{i,j-1})$ moved outside the integral. This remains true when we would enhance it with additional risk factors, earlier values from the longitudinal data or include the treatment indicator in the hazard function. It should be noted that it is more realistic to have the treatment model depend on $Y_{ij}$ instead of $\bm{r}_i$ since treatment interventions would be based on the observations rather than the expected observations. If we further assume that the treatment model does not share any common unknown parameters with the longitudinal and time-to-event models (ignorability), the unknown parameters in the treatment model can be ignored. Therefore, to obtain the estimates of the unknown parameters in the longitudinal and time-to-event models, the following partial likelihood for individual $i$ is considered:
\begin{equation*}
    \begin{split}
        L^{{\mathrm{partial}}}_i = \int &f(Y_{i1}|\bm{r}_i)\prod_{j=2}^{m_i} f(Y_{ij}|Z_{i,j-1},\bm{r}_i)f(T_i,D_i|\bm{r}_i) f(\bm{r}_i)d\bm{r}_i\\
         = \int &\prod_{j=1}^{m_i} \phi\left(\frac{y_{ij}-\mu_{ij}}{\sigma_{ij}}\right)  \phi\left(\sqrt{(\bm{r}_i-\bm{\theta})' \bm{\Sigma}^{-1}(\bm{r}_i-\bm{\theta})}\right)\\
        &\times \lambda_i^{D_i}(T_i)\exp\left(-\int_0^{T_i}\lambda_i(s)ds\right)d\bm{r}_i,
    \end{split}
\end{equation*}
where $\phi(\cdot)$ is the standard normal density function. Furthermore, the integration of the hazard function with respect to time in the time-to-event model is approximated by

\begin{align*}
\int_0^{T_i} \lambda_i(s) ds =& \sum_{j=1}^{m_i} \int_{t_{i,j-1}}^{t_{ij}}\lambda_i(s) ds  + \int_{t_{i,m_i}}^{T_i}\lambda_i(s) ds\\
\approx& \sum_{j=1}^{m_i} (\Lambda_0(t_{ij}) - \Lambda_0(t_{i,j-1}))\exp(\gamma_0 \mu_{i,j-1} + \gamma_1 \log \sigma^2_{i, j-1})\\
& + (\Lambda_0(T_i)-\Lambda_0(t_{i,m_i}))\exp(\gamma_0 \mu_{i,m_i} + \gamma_1 \log \sigma^2_{i,m_i})
\end{align*}
using an event history formulation similar to Henderson~\cite{Henderson2000}, where $\Lambda_0$ is the cumulative baseline hazard function. We implicitly assumed a piece-wise constant function for both $\mu_{ij}$ and $\sigma_{ij}$ within the interval $[t_{ij}, t_{i,j+1})$ that is left-continuous with respect to time.

Unfortunately, direct maximization of the partial likelihood may yield biased estimates for $\gamma_1$ due to the effect of the variability in the hazard function. The reason is that the maximum likelihood estimator tends to underestimate the variance components of the longitudinal profile in small sample cases is due to the failure to account for a reduction in degrees of freedom associated with fixed effects parameters. The underestimated variance component of the latent variable $c_i$ will consequently lead to a underestimated $\sigma_{ij}^2$, which in turn will cause overestimation of $\gamma_1$ in the time-to-event model. Therefore, a restricted maximum likelihood (REML) estimator is needed to transform the partial likelihood such that the distribution of the error construct is unrelated to those parameters. However, finding a transformation of the data consisted of both longitudinal profile and time-to-event outcome is not trivial. One alternative can be derived using a Bayesian formulation by introducing a flat prior for the fixed effect parameters. This approach was considered by Harville~\cite{Harville1974, Harville1976} and Dempster et al.~\cite{Dempster1981} and has been shown to coincide with the REML estimates. Furthermore, the empirical Bayes estimates of the latent variables are the estimated means of the posterior distributions~\cite{Laird1982}. Therefore, we take a Bayesian approach to estimate the parameters of interests by introducing prior distributions to the fixed effect parameters.

\section{Sampling posterior distribution via MCMC}\label{sec:mcmc}
Unbounded uniform priors on $(-\infty, \infty)$ are specified for all fixed effects parameters $\beta_0$, $\beta_1$, $\beta_2$, $\beta_3$, $\nu$, $\gamma_0$, $\gamma_1$ independent of each other. A truncated half-normal prior was assigned to $\sigma_0$. For the parameters involved in the baseline survival distribution (e.g., Weibull parameters), bounded uniform priors are specified. For instance, for the shape and scale parameters of the Weibull baseline, we considered a $[0, U]$-uniform distribution with $U >> \hat{U}$. Here $\hat{U}$ is the estimated parameter value obtained from a standalone fit of the time-to-event model to the data. For the covariance matrix we assume an uniform prior over the set of all symmetric, positive-definite matrices. % all positive-defitine matrices.

%For the covariance matrix $\bm{\Sigma}$, we decompose the prior into a scale and a correlation matrix. Specifically, $\bm{\Sigma} = \mathrm{diag}(\bm{\tau})\, \bm{\Omega}\, \mathrm{diag}(\bm{\tau})$, where $\bm{\tau}$ is a vector of scale coefficients $\tau_k = \sqrt{\bm{\Sigma}_{k,k}}$ of the $(k,k)$-diagonal of the covariance matrix, $\bm{\Omega}$ the correlation matrix with its $(p,q)$ entry $\Omega_{p,q} = (\tau_p \tau_q)^{-1} \bm{\Sigma}_{p,q}$. A half-Cauchy distribution $C^{+}(0, 2.5)$ is specified as the prior for $\tau_k$, and a LKJ correlation distribution~\cite{lkj} is specified for the correlation matrix $\bm{\Omega}$.\cite{Gelman2013} The primary motivation for adopting this weakly-informative prior distribution for $\tau_k$ is to constrain the intra-individual variance before and after treatment away from very large values in case of insufficient number of follow-up measurements for some individuals~\cite{Gelman2006}. Further re-parametrization of $\bm{\tau}$ and Cholesky factorization of the correlation matrix $\bm{\Omega}$ are used only for computational purposes. Posterior samples are generated using the No U-Turn Sampler (NUTS), an extension to the Hamiltonian Monte Carlo algorithm~\cite{nuts}.

\section{Simulations}\label{sec:sim}
\subsection{Data generation}
The simulated data is generated according to the three models described in Section~\ref{sec:method}. For each individual, the latent variables $\bm{r}_i$ were drawn from a multivariate normal distribution $\mathcal{N}(\bm{\theta}, \bm{\Sigma})$, where $\bm{\theta} = (12, 0.1, -0.3, -0.05, 0)'$ and
\begin{equation*}
\bm{\Sigma} = 
    \begin{pmatrix*}[r]
            4&  -0.02&   -0.2&  -0.05&    0.6\\
        -0.02&  0.005&  -0.015& 0.005&   0.01\\
         -0.2&  -0.015&      1&  -0.02&   -0.1\\
        -0.05&  0.005&  -0.02&  0.015& -0.025\\
         0.6 &   0.01&   -0.1& -0.025&    0.3\\
    \end{pmatrix*}.
\end{equation*}
Furthermore, the $j$th ($j=1,\dots, m_i=30)$ measurement time was generated by drawing uniformly from the interval $[j, j+1)$. Afterwards, the longitudinal risk factor profile and the treatment (of the risk factor) indicator $Y_{ij}, Z_{ij}$ were generated recursively according to $Z_{i,j-1}$ and $Y_{i, j-1}$ following the longitudinal and the treatment model, respectively. Treatment at baseline $Z_{i0}$ is taken to be 0 for all individuals. For the treatment model, the corresponding parameters were specified to be $\alpha_0 = -3$ and $\alpha_1 = -0.05$ such that $30\%$ of the individuals with 30 repeats received treatment during the course of the follow-up. With double the amount of repeats, the percentage of people receiving treatment went up to $45\%$. While, in the longitudinal profile, $\sigma_0 = 2$, and $\nu = 0.5$. For the time-to-event model, a Weibull baseline hazard function 
\begin{equation*}
    \lambda_0(t) = \frac{k}{\xi} \left(\frac{t}{\xi}\right)^{k-1}
\end{equation*}
was specified with the scale parameter $\xi = 150$ and shape parameter $k = 1.5$. Event time $T_i$ was determined using the previously generated conditional mean of the longitudinal profile and the intra-individual variability according to the time-to-event model. The corresponding coefficients were set to $\gamma_0 = 0.05$ and $\gamma_1 = 0.2$. Since both the conditional mean of the longitudinal profile and the intra-individual variability are time-dependent covariates, a simulation algorithm proposed by Prikken et al.~\cite{Prikken} for generating event time according to the Cox's regression model with time-dependent covariates was used. After the event time was determined, data related to observations at periods later than the event time were discarded. In case the simulated event time is larger than the last measurement time of the longitudinal profile, the individual was considered to be censored at the last time point of the longitudinal data.

In total 500 simulation runs were performed with each simulation containing 500 individuals. The proposed model was fitted to each simulated data set using the Stan Bayesian statistical analysis and computation platform~\cite{stan}. For each simulation run, 2 chains of 1000 posterior samples per chain were generated by the MCMC algorithm with the first $2\times500$ samples per chain as warm-up. We also verified that the MCMC chains converged well, using the R-hat statistic. The posterior means of the generated posterior samples were used as the point estimates of the unknown parameters for each simulation run. The mean, empirical standard error (SD) and the mean squared error (MSE) were calculated based on the 500 point estimates for each unknown parameters. Two extra simulations were run, two with a higher number of repeats, $m_i\in\{45,60\}$, and one with a higher number of individuals $n=750$.

\subsection{Results}
The results of the simulation with $m_i=30$ are shown in Table~\ref{tab:sim_results1} and \ref{tab:sim_results2}, the results for the three extra simulations can be found in the Supplemental material in Tables \ref{tab:sim_results3} till \ref{tab:sim_results8}. In Table~\ref{tab:sim_results1}, the mean, SD, and MSE of the fixed effect were shown. All parameters were estimated without bias, with the exemption of the baseline parameters for the hazard function. Empirical standard errors for $\beta_0$ and $\beta_2$ were relatively large and the coverage probabilities of $\beta_0$ was somewhat conservative (i.e., below the nominal value) but still within the range of simulation (Monte Carlo) variability. The coverage of the 'k' parameter of the baseline hazard is liberal. As the number of repeats or simulations increased in the extra simulations the bias in the baseline hazard decreased.In Table~\ref{tab:sim_results2}, the average estimates and the corresponding empirical standard error of the covariance matrix $\bm{\Sigma}$ is displayed. Compared to the true value used in the simulation, covariance estimates for the variance of the slope after treatment ($\mathrm{var}(\beta_3)$) was overestimated and the covariance between the slope and intercept change after treatment ($\mathrm{cov}(\beta_1, \beta_3)$) was underestimated. The bias in the covariance matrix estimation did not  affect the fixed effects in the longitudinal profile and the hazard function. Therefore, this approach provides a REML-like estimates of the fixed effects that are independent of the variance parameters.

\begin{table}
\small\sf\centering
\caption{Simulation results: mean, empirical standard error (SD), the mean squared error (MSE), and the coverage probability (CP) of the fixed effect parameters and of the $\tau_k$ in both the longitudinal and survival model based on 500 simulations, 100 individuals and 30 measurements.}\label{tab:sim_results1}
\begin{tabular}{lrcccc}
\hline
& & \multicolumn{3}{c}{\textbf{Point estimate}}\\ \cline{3-6} 
\textbf{Parms} & \textbf{True value} & \textbf{Mean} & \textbf{SD} & \textbf{MSE} & \textbf{CP}\\ \hline
\multicolumn{6}{l}{\textit{Longitudinal model}}\\ \hline
$\beta_0$      & 12     & 11.9943 & 0.2103 & 0.0442 & 94.6\% \\
$\beta_1$      & 0.0989 & 0.0136 & 0.0002 & 96.6\% \\
$\beta_2$      & -0.3   & -0.3025 & 0.0744 & 0.0055 & 94.6\% \\
$\beta_3$      & -0.05  & -0.0483 & 0.0227 & 0.0005 & 95.8\% \\
$\nu$          & 0.5    & 0.5048 & 0.0735 & 0.0054 & 94.6\% \\
$\sigma_0$     & 1.414  & 1.4149 & 0.0542 & 0.0029 & 94.6\% \\\hline
\multicolumn{6}{l}{\textit{Time-to-event model}}\\ \hline
$\gamma_0$      & 0.05  & 0.0771 & 0.1112 & 0.0131 & 91.8\% \\
$\gamma_1$      & 0.2   & 0.2265 & 0.6905 & 0.4765 & 93.4\% \\
$k$             & 1.5   & 1.1211 & 0.1199 & 0.1579 & 95.0\% \\
$-k\log(\xi)$   &-7.516 & -6.9931 & 1.2166 & 1.7506 & 98.2\% \\ \hline
\end{tabular}
\end{table}

\begin{table}
  \small\sf\centering
\caption{Simulation results: mean (and empirical standard error) and the coverage of the covariance matrix $\bm{\Omega}$ based on 500 simulations, 500 individuals and 30 measurements.}\label{tab:sim_results2}
\begin{tabular}{@{}cccccc@{}}
\toprule
       & $b_{0i}$    & $b_{1i}$   & $b_{2i}$    & $b_{3i}$   & $c_i$      \\ \midrule
$b_{0i}$   & 4.0211 (0.6574)  & -0.0232 (0.0189)  & -0.1719 (0.2483)  & -0.0325 (0.0318)  & 0.5387 (0.1324) \\ 
$b_{1i}$   & -  & 0.0060 (0.0019)  & -0.0064 (0.01)  & 0.0015 (0.0012)  & 0.0049 (0.0051)  \\ 
$b_{2i}$   & -  & - &  0.9618 (0.3504)  & -0.0127 (0.0187)  & -0.0626 (0.0688) \\ 
$b_{3i}$   & -  & - & - & 0.0216 (0.0059)  & -0.0156 (0.0097) \\ 
$c_i$      & -  & - & - & - & 0.3060 (0.0555) \\ \bottomrule
\toprule
       & $b_{0i}$    & $b_{1i}$   & $b_{2i}$    & $b_{3i}$   & $c_i$      \\ \midrule
$b_{0i}$   & 94.2\%& 99.0\%& 97.4\%& 96.6\%& 91.2\% \\
$b_{1i}$   & -  &  92.8\%& 96.2\%& 57.4\%& 93.8\% \\
$b_{2i}$   & -  & - & 92.0\%& 98.6\%& 96.0\% \\ 
$b_{3i}$   & -  & - & - &84.6\%& 91.4\% \\
$c_i$      & -  & - & - & - & 95.2\% \\ \bottomrule
\end{tabular}
\end{table}

\section{Case study}\label{sec:fhs}
The proposed joint model was fitted to the Framingham Heart Study data described in Section~\ref{sec: fhs_eda}. The longitudinal profile considered was the SBP, and the event of interests was the occurrence of CVD. The objective was to investigate the effect of initiation and maintenance of anti-hypertensive medication on the SBP profile and the variability of the SBP, the indirect effect of the treatment on the risk of developing CVD, and the effect of intra-individual variability on the risk of CVD. Time to CVD event was considered to be the time of clinical diagnosis. Though an interval-censored time-to-event model would be more appropriate, the approximation of the true event time by the diagnosis time at the total time scale suffice to illustrate the usage of the proposed joint model. Furthermore, a Weibull baseline distribution was assumed for the time-to-event model based on the recommendation from the literature~\cite{Hu2013}. All posterior estimates of the model generated from the MCMC algorithm converged well with the split R-hat statistic (potential scale reduction factor)~\cite{Gelman1992} close to 1 for all parameters. Additional graphic inspection of the trace-plots was also performed. Posterior predictive checking~\cite{Gelman1996} was performed to assess the goodness-of-fit of the proposed model to the Framingham Heart Study data and did not indicate lack of fits. (The results of the posterior predictive checking are provided in Supplementary material).

The results of the estimates of the fixed effect parameters is summarized in Table~\ref{tab:fhs_results1}. For the longitudinal model, it can be seen that the average SBP at baseline was estimated to be 129 mmHg with an annual natural (without anti-hypertensive medication) growth rate of 0.74 mmHg/year. The anti-hypertensive medication is estimated to reduce the absolute value of the SBP by 2.64 mmHg at initiation of the treatment on average and the growth rate by 0.52 mmHg/year. The main interest, however, lies in the effect of antihypertensive treatment on the variability of the SBP. The parameter of the effect of treatment on the variability was estimated to be 0.56, which corresponds to a $1.75 \approx \exp(0.56)$ times higher SBP variability after treatment. On the other hand, the parameter estimates in the time-to-event model indicate that a 10 mmHg increase in the SBP is associated with a hazard ratio (for CVD) of $\exp(0.15) \approx 1.16$. Furthermore, 2 times higher SBP standard deviation is estimated to have a hazard ratio of $\exp(2\times 0.38\times \log 2) \approx 1.69$. Thus, whereas the use of antihypertensive treatment reduces the risk on CVD through the lowering of SBP, it may potentially increase the risk of CVD by increasing the variability of SBP. Since the effect of the change in variability (of SBP) is larger than the effect of the change in SBP, our results are in line with the finding that patients receiving anti-hypertensive medication still appear to have higher risk of CVD than those not on anti-hypertensive medication with the same SBP level~\cite{Agosotino2008, Psaty2001, Chambless2003, Chambless2004}. Furthermore, individuals that will immediately benefit from the anti-hypertensive medication are those whose subject-specific reduction of the SBP $b_{2i}$ satisfies the inequality $\gamma_0 b_{2i} + \gamma_1 \nu \le 0$. Since $b_{2i}$ was estimated to have mean $-2.640$ and variance $111.41$, the percentage of individuals with immediate benefits from the treatment is approximately $\Phi((-13.851 + 2.640)/\sqrt{111.41}) \approx 14.41\%$.

\begin{table}
\small\sf\centering
\caption{Fixed effect parameter estimates of the joint model fitted to the Framingham Heart Study data}\label{tab:fhs_results1}
\begin{tabular}{@{}lccccc@{}}
\toprule
\textbf{Parms} & \textbf{\begin{tabular}[c]{@{}c@{}}Posterior \\Mean\end{tabular}}& \textbf{SD}  & \multicolumn{2}{c}{\textbf{95\% Quantile Interval}}& \textbf{MCSE} \\ \midrule
\multicolumn{5}{l}{\textit{Longitudinal model}}                                                                                                            \\ 
$\beta_0$      & 128.963 & 0.297   & 128.371  & 129.530   & 0.016\\
$\beta_1$      & 0.741   & 0.014   & 0.714    & 0.768     & 0.0004        \\
$\beta_2$      & -2.640  & 0.448   & -3.53    & -1.758    & 0.011   \\
$\beta_3$      &-0.516   & 0.049   & -0.615   & -0.421    & 0.002\\
$\nu$          & 0.5567  & 0.0215  & 0.5425   & 0.5988    & 0.0005 \\
$\sigma_0$     & 11.524  & 0.065   & 11.397   & 11.652    & 0.002     \\ \midrule
\multicolumn{5}{l}{\textit{Survival model}}                                                                                                             \\ 
$\gamma_0$     & 0.01539 & 0.00062 & 0.0141   & 0.0165    & 0.00001  \\
$\gamma_1$     & 0.3829  & 0.0494  & 0.2858   & 0.4167    & 0.0011   \\ \bottomrule
\end{tabular}
\end{table}

The estimated correlation matrix (posterior mean) of the five latent variables is presented in Table~\ref{tab:fhs_results2}. A small negative correlation is found between the baseline SBP level ($b_{0i}$) and the slope prior to treatment ($b_{1i}$) which indicates that individuals with higher baseline SBP experience a slower natural growth rate of SBP. It is however more interesting to see that the two effects of the treatment, namely the direct lowering effect on the SBP level ($b_{2i}$) and the reduction of the growth rate of the SBP ($b_{3i}$), are all negatively correlated with the baseline SBP and the growth rate prior to the treatment. Heuristically, this means that individuals with higher baseline SBP and faster SBP increase before treatment are more likely to benefit from the anti-hypertensive medication either by a direct SBP lowering and/or by a larger reduction in the progression after treatment. Furthermore, the correlation between SBP variability and SBP level as mentioned in Section~\ref{sec:intro}, is captured by the large positive correlation between $b_{0i}$ and $c_i$ in Table~\ref{tab:fhs_results2}. Not surprisingly, a higher SBP level is associated with a higher variability and the SBP variability is positively correlated with the growth rate. However, more strikingly, both treatment effects are negatively correlated with the SBP variability, indicating that the anti-hypertensive medication is more beneficial for individuals with a higher SBP variability. Since higher variability corresponds to higher SBP levels and faster progression before treatment, this finding is consistent with the previously presented findings.

\begin{table}
\small\sf\centering
\caption{Estimated correlation matrix of the random effect of the joint model fitted to the Framingham Heart Study data}\label{tab:fhs_results2}
\begin{tabular}{@{}l|ccccc@{}}
\toprule
         & $b_{0i}$    & $b_{1i}$   & $b_{2i}$    & $b_{3i}$   & $c_i$      \\ \midrule
$b_{0i}$ & 1.000  & -0.1763 & -0.0986 & -0.2360 & 0.6285  \\
$b_{1i}$ & -0.1763 & 0.9991  & -0.2782 & -0.5728 & 0.2691  \\
$b_{2i}$ & -0.0986 & -0.2782 & 0.9945  & -0.1319 & -0.1241 \\
$b_{3i}$ & -0.2360 & -0.5728 & -0.1319 & 0.9926  & -0.3713 \\
$c_i$    & 0.6285  & 0.2691  & -0.1241 & -0.3713 & 0.9954  \\ \bottomrule
\end{tabular}
\end{table}

\section{Discussion}\label{sec:discussion}
In this investigation, we have formulated an innovative joint model for longitudinal risk factor profiles and event outcome. It incorporates direct modeling of the intra-individual variability of the longitudinal profile in both the longitudinal and time-to-event model and it handles the impact of disruptive patterns in the longitudinal profile of the putative risk factor. This model provides opportunities to investigate the effect of treatment on both the longitudinal profile of a given risk factor and its residual variance, and relate these changes to the risk of outcome events on follow-up. Simulation results showed that the proposed method was able to estimate all fixed effect parameters in both the longitudinal and time-to-event model (almost) without bias. To obtain numerical stability, we decomposed the variance-covariance matrix of all random effects into a scale and a correlation matrix and further decomposed the correlation matrix using Cholesky decomposition. However, for 100 individuals per simulation, sample size was not sufficient to guarantee the convergence to the true values of the parameters in the correlation matrix. A possible improvement is the use of Fisher's z transformation on the correlation parameters to make the distribution of the estimates closer to a normal distribution at lower numbers of individuals. Nevertheless, for the analysis of the Framingham Heart Study this transformation was not needed due to much larger sample size. In the analysis of the Framingham Heart Study using the proposed joint model, we were able to identify the pros and cons of the anti-hypertensive medication. It reduced the absolute value of the SBP and decreases its progression but at the expense of a higher SBP variability. Since our model demonstrated that higher variability is associated with elevated risk of developing CVD, the beneficial effects of treatment on lowering SBP may be partly or possibly fully offset by an increase in variability, a premise worthy of investigating in the future. 

It should be noted that the parameters in the treatment allocation model were not discussed in the present investigation since the treatment allocation model factorizes the likelihood function and it can be ignored. However, it is straightforward to incorporate the treatment allocation model by considering the complete likelihood function instead of the partial likelihood function.

There are several ways of improving the treatment model without changing the partial likelihood of the joint model. This includes adding earlier longitudinal measurements or treatment indicators, covariates, and even the possibility of cessation of the treatment. In this case, the cessation of the treatment can only depend on the past longitudinal observations or treatment indicators. As long as the treatment probability does not depend on the random variable of the joint model, the partial likelihood is as determined in Section \ref{sec:method}. Note that if cessation is added to the treatment model, while the longitudinal and time-to-event model stay the same, cessation of the treatment will have the same result at that time as if the treatment was never started. For treatment to have effect even if it is stopped the longitudinal model will need to be adjusted.

Though a Bayesian framework is adopted, other estimation methods such as the expectation maximization (EM) algorithm and its variations, for instance Monte Carlo EM~\cite{Wei1990} and PX-EM~\cite{Liu1998}, would also be possible. Furthermore, the proposed joint model is flexible in the way that the modeling of the longitudinal profile and event outcome together with the modeling of the longitudinal profile variability can include additional covariates to reflect the domain-specific knowledge of the phenomenon that one tries to investigate. For instance, in the Framingham Heart Study or other studies of CVD, intervenable risk factors such as smoking behavior of the individual and its effect on both the blood pressure profile and the risk of developing CVD may be of interests to the investigators. Then inclusion of this additional time-dependent covariate similar to the time-dependent treatment in the current joint model may be introduced.

Finally, in the current joint model the association between the longitudinal profile and the event outcome is linked directly via the time-dependent conditional mean and variance of the longitudinal profile. However, one may wish to adopt the shared latent variable joint model framework, which uses the latent variables $\bm{\theta}$ themselves in the time-to-event model. This may be of interests for future studies as well.

\begin{sm}
To check the fit of the model to the Framingham Heart Study data, 1000 New observations $y_{\mathrm{rep}}$ was drawn from the posterior predictive distribution of the longitudinal profile $y$ per observation in the Framingham Heart Study data set (see Figure~\ref{fig:ppc}). 
\begin{figure}[!htp]
    \centering
    \includegraphics[width=0.6\textwidth]{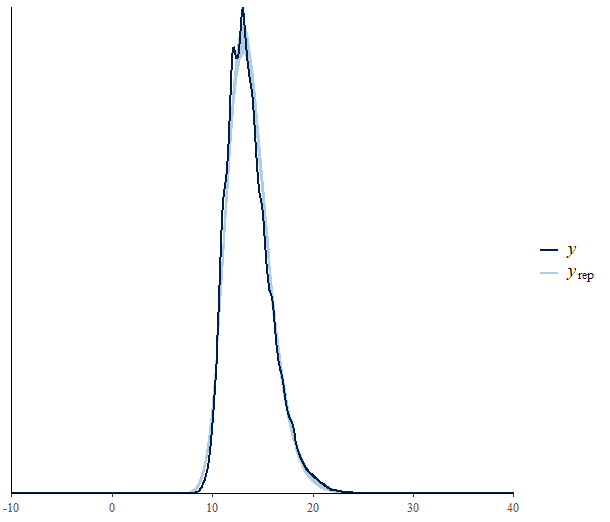}
    \caption{Kernel density estimates of yrep overlaid with the distribution of y itself (The scale is $1/10$ of the original SBP)}
    \label{fig:ppc}
\end{figure}

\begin{table}
\small\sf\centering
\caption{Simulation results: mean, empirical standard error (SD), the mean squared error (MSE), and the coverage probability (CP) of the fixed effect parameters and of the $\tau_k$ in both the longitudinal and survival model based on 500 simulations, 100 individuals and 60 measurements.}\label{tab:sim_results3}
\begin{tabular}{lrcccc}
\hline
& & \multicolumn{3}{c}{\textbf{Point estimate}}\\ \cline{3-6} 
\textbf{Parms} & \textbf{True value} & \textbf{Mean} & \textbf{SD} & \textbf{MSE} & \textbf{CP}\\ \hline
\multicolumn{6}{l}{\textit{Longitudinal model}}\\ \hline
$\beta_0$      & 12     & 11.9937 & 0.2066 & 0.0427 & 94.4\% \\
$\beta_1$      & 0.1    & 0.0991 & 0.0120 & 0.0001 & 96.4\% \\
$\beta_2$      & -0.3   & -0.3002 & 0.0657 & 0.0043 & 94.8\% \\
$\beta_3$      & -0.05  & -0.0485 & 0.0178 & 0.0003 & 96.4\% \\
$\nu$          & 0.5    & 0.5048 & 0.0576 & 0.0033 & 95.2\% \\
$\sigma_0$     & 1.414  & 1.4146 & 0.0522 & 0.0027 & 94.8\% \\\hline
\multicolumn{6}{l}{\textit{Time-to-event model}}\\ \hline
$\gamma_0$      & 0.05  & 0.0585 & 0.0278 & 0.0008 & 95.8\% \\
$\gamma_1$      & 0.2   & 0.2218 & 0.3016 & 0.0913 & 93.4\% \\
$k$             & 1.5   & 1.4904 & 0.0916 & 0.0085 & 100.0\% \\ 
$-k\log(\xi)$   &-7.516 & -7.6640 & 0.3612 & 0.1522 & 100.0\% \\ \hline
\end{tabular}
\end{table}

\begin{table}
  \small\sf\centering
\caption{Simulation results: mean (and empirical standard error) and the coverage of the covariance matrix $\bm{\Omega}$ based on 500 simulations, 100 individuals and 60 measurements.}\label{tab:sim_results4}
\begin{tabular}{@{}cccccc@{}}
\toprule
       & $b_{0i}$    & $b_{1i}$   & $b_{2i}$    & $b_{3i}$   & $c_i$      \\ \midrule
$b_{0i}$   & 4.0000 (0.6551)  & -0.0227 (0.0184)  & -0.1673 (0.2475)  & -0.0348 (0.0278)  & 0.5394 (0.1319)\\
$b_{1i}$   & -  & 0.0059 (0.0017)  & -0.0105 (0.01)  & 0.0028 (0.0012)  & 0.0065 (0.0052)\\
$b_{2i}$   & -  & - &  1.0009 (0.2688)  & -0.0126 (0.017)  & -0.0669 (0.0648)\\
$b_{3i}$   & -  & - & - & 0.0185 (0.0039)  & -0.0186 (0.0084)\\
$c_i$      & -  & - & - & - & 0.3033 (0.0521)\\ \bottomrule
\toprule
       & $b_{0i}$    & $b_{1i}$   & $b_{2i}$    & $b_{3i}$   & $c_i$      \\ \midrule
$b_{0i}$   &  94.2\%& 98.6\%& 96.0\%& 96.4\%& 90.2\%\\
$b_{1i}$   & -  & 95.2\%& 98.0\%& 73.8\%& 94.6\%\\
$b_{2i}$   & -  & - & 94.2\%& 95.8\%& 95.4\%\\
$b_{3i}$   & -  & - & - &90.4\%& 91.6\%\\
$c_i$      & -  & - & - & - &  94.6\%
\\ \bottomrule
\end{tabular}
\end{table}
\begin{table}
\small\sf\centering
\caption{Simulation results: mean, empirical standard error (SD), the mean squared error (MSE), and the coverage probability (CP) of the fixed effect parameters and of the $\tau_k$ in both the longitudinal and survival model based on 500 simulations, 500 individuals and 30 measurements.}\label{tab:sim_results5}
\begin{tabular}{lrcccc}
\hline
& & \multicolumn{3}{c}{\textbf{Point estimate}}\\ \cline{3-6} 
\textbf{Parms} & \textbf{True value} & \textbf{Mean} & \textbf{SD} & \textbf{MSE} & \textbf{CP}\\ \hline
\multicolumn{6}{l}{\textit{Longitudinal model}}\\ \hline
$\beta_0$      & 12     & 12.004 & 0.0964 & 0.0093 & 94 \% \\
$\beta_1$      & 0.1    &  0.0996 & 0.0063 & 0.0000 & 94.8\% \\
$\beta_2$      & -0.3   & -0.2935 & 0.0749 & 0.0056 & 94.4\% \\
$\beta_3$      & -0.05  & -0.0494 & 0.01 & 0.0001 & 95.2\% \\
$\nu$          & 0.5    & 0.5054 & 0.0316 & 0.001 & 94.2\% \\
$\sigma_0$     & 1.414  & 1.4126 & 0.0236 & 0.0006 & 95\% \\\hline
\multicolumn{6}{l}{\textit{Time-to-event model}}\\ \hline
$\gamma_0$      & 0.05  &  0.0583 & 0.0363 & 0.0014 & 95.4\% \\
$\gamma_1$      & 0.2   &  0.2227 & 0.2095 & 0.0443 & 95.2\% \\
$k$             & 1.5   & 1.6487 & 0.0744 & 0.0277 & 99.4\%\\ 
$-k\log(\xi)$   &-7.516 & -8.1792 & 0.3826 & 0.586 & 95.2\% \\ \hline
\end{tabular}
\end{table}

\begin{table}
  \small\sf\centering
\caption{Simulation results: mean (and empirical standard error) and the coverage of the covariance matrix $\bm{\Omega}$ based on 500 simulations, 500 individuals and 60 measurements.}\label{tab:sim_results6}
\begin{tabular}{@{}cccccc@{}}
\toprule
       & $b_{0i}$    & $b_{1i}$   & $b_{2i}$    & $b_{3i}$   & $c_i$      \\ \midrule
$b_{0i}$   & 3.9995 (0.2989)  & -0.0215 (0.0121)  & -0.2 (0.1369)  & -0.0426 (0.0189)  & 0.5891 (0.0660)\\
$b_{1i}$   & -  & 0.0053 (0.0008)  & -0.0108 (0.0077)  & 0.0034 (0.0009)  & 0.0081 (0.0031) \\ 
$b_{2i}$   & -  & - & 0.973 (0.1489)  & -0.0175 (0.0126)  & -0.0904 (0.0402) \\ 
$b_{3i}$   & -  & - & - & 0.0178 (0.0024)  & -0.0217 (0.0055) \\
$c_i$      & -  & - & - & - &0.3016 (0.0249) \\  \bottomrule
\toprule
       & $b_{0i}$    & $b_{1i}$   & $b_{2i}$    & $b_{3i}$   & $c_i$      \\ \midrule
$b_{0i}$   &  94.8\%& 93.8\%& 96.6\%& 94.6\%& 94.0\% \\ 
$b_{1i}$   & -  & 92.6\%& 94.4\%& 66.4\%& 93.4\% \\ 
$b_{2i}$   & -  & - & 94.8\%& 95.4\%& 97.0\% \\ 
$b_{3i}$   & -  & - & - &85.4\%& 92.0\% \\ 
$c_i$      & -  & - & - & - &94.8\% \\ 
\\ \bottomrule
\end{tabular}
\end{table}
\begin{table}
\small\sf\centering
\caption{Simulation results: mean, empirical standard error (SD), the mean squared error (MSE), and the coverage probability (CP) of the fixed effect parameters and of the $\tau_k$ in both the longitudinal and survival model based on 500 simulations, 750 individuals and 30 measurements.}\label{tab:sim_results7}
\begin{tabular}{lrcccc}
\hline
& & \multicolumn{3}{c}{\textbf{Point estimate}}\\ \cline{3-6} 
\textbf{Parms} & \textbf{True value} & \textbf{Mean} & \textbf{SD} & \textbf{MSE} & \textbf{CP}\\ \hline
\multicolumn{6}{l}{\textit{Longitudinal model}}\\ \hline
$\beta_0$      & 12     & 12.0015 & 0.0766 & 0.0059 & 95.0\% \\
$\beta_1$      & 0.1    & 0.1001 & 0.0049 & 0.00002 & 94.6\% \\
$\beta_2$      & -0.3   & -0.3030 & 0.0603 & 0.0036 & 94.8\% \\
$\beta_3$      & -0.05  & -0.0499 & 0.0083 & 0.0001 & 95.8\% \\
$\nu$          & 0.5    & 0.4994 & 0.0258 & 0.0007 & 94.2\% \\
$\sigma_0$     & 1.414  & 1.4145 & 0.0195 & 0.0004 & 95.2\% \\\hline
\multicolumn{6}{l}{\textit{Time-to-event model}}\\ \hline
$\gamma_0$      & 0.05  & 0.0459 & 0.0292 & 0.0009 & 94.6\% \\
$\gamma_1$      & 0.2   & 0.2063 & 0.1667 & 0.0278 & 94.2\% \\
$k$             & 1.5   & 1.4979 & 0.0466 & 0.0022 & 100.0\% \\ 
$-k\log(\xi)$   &-7.516 & -7.3110 & 0.3020 & 0.1330 & 99.8 \% \\ \hline
\end{tabular}
\end{table}

\begin{table}
  \small\sf\centering
\caption{Simulation results: mean (and empirical standard error) and the coverage of the covariance matrix $\bm{\Omega}$ based on 500 simulations, 750 individuals and 30 measurements.}\label{tab:sim_results8}
\begin{tabular}{@{}cccccc@{}}
\toprule
       & $b_{0i}$    & $b_{1i}$   & $b_{2i}$    & $b_{3i}$   & $c_i$      \\ \midrule
$b_{0i}$   & 4.0732 (0.2439) & -0.0217 (0.0095) & -0.2047 (0.1158) & -0.0486 (0.0161) & 0.6091 (0.0512)\\
$b_{1i}$   & -  & 0.0053 (0.0004) & -0.0151 (0.0073) & 0.0045 (0.0009) & 0.01 (0.0029)\\
$b_{2i}$   & -  & - & 1.0352 (0.1205) & -0.02 (0.0111) & -0.1012 (0.0343)\\
$b_{3i}$   & -  & - & - & 0.0163 (0.002) & -0.025 (0.0048)\\
$c_i$      & -  & - & - & - &  0.3065 (0.0193) \\\bottomrule
\toprule
       & $b_{0i}$    & $b_{1i}$   & $b_{2i}$    & $b_{3i}$   & $c_i$      \\ \midrule
$b_{0i}$   & 92.0\% & 95.6\% & 96.6\% & 95.6\% & 95.4\%\\
$b_{1i}$   & -  &  93.4\% & 94.2\% & 90.8\% & 94.8\%\\
$b_{2i}$   & -  & - & 94.2\% & 94.4\% & 96.2\%\\
$b_{3i}$   & -  & - & - &91.4\% & 95.8\%\\
$c_i$      & -  & - & - & - & 95.0\% \\ \bottomrule
\end{tabular}
\end{table}

\end{sm}

\begin{acks}
We acknowledge the dedication of the Framingham Heart Study participants without whom this research would not be possible.
\end{acks}

\begin{dci}
The author(s) declared no potential conflicts of interest with respect to the research, authorship, and/or publication of this article.
\end{dci}

\begin{funding}
This work was supported by the ITEA3 [15032] to Z. Zhan and E.R. van den Heuvel;
by the Evans Medical Foundation and the Jay and Louis Coffman Endowment from the Department of Medicine, Boston University School of Medicine to V. Ramachandran. The Framingham Heart Study was supported by the National Heart, Lung and Blood institute [NO1-HC-25195, HHSN268201500001I, 75N92019D00031].
\end{funding}

\end{document}